\documentclass[fleqn,twoside]{article}
\usepackage[headings]{espcrc2}

\readRCS
$Id: espcrc2.tex,v 1.2 2004/02/24 11:22:11 spepping Exp $
\usepackage{graphicx}
\usepackage[figuresright]{rotating}

\newcommand{\AmS}{{\protect\the\textfont2
  A\kern-.1667em\lower.5ex\hbox{M}\kern-.125emS}}

\title{Consequences of the common origin of the knee and
ankle in Cosmic Ray Physics}

\author{Antonio Codino\address[PG]{INFN and Dipartimento di Fisica dell'Universit\`a  di Perugia, Italy.}
        and
        Fran\c{c}ois Plouin\address[FRD]{Former CNRS Researcher Ecole Polytechnique, LLR, F-91128 Palaiseau, France.}}

\begin{document}

\begin{abstract}
The differential energy spectrum of the cosmic radiation from solar
modulation energies up to $5\times10^{19}$ eV is correctly predicted
by a recent theory of the knee and ankle which uses only one
normalization point. This remarkable quantitative result, spanning
over many decades in energy and intensity, along with the existence
of the second knee at $6\times10^{17}$ eV, is obtained assuming
constant spectral indices of individual ions at the cosmic-ray
sources and no other critical hypotheses. In this study the chemical
composition of the cosmic radiation is evaluated as a direct
consequence of the theory. The computed mean logarithmic mass
exhibits a rising trend from $1.8$ to $3.0$ in the range
$10^{15}$-$10^{17}$ eV, a maximum value of $3.2$ at $3 \times
10^{17}$ eV, and a characteristic lightening above $3 \times
10^{17}$ eV up to $4 \times 10^{18}$ eV. All of these distinctive
features are in accord with the data of many experiments. Two
additional consequences intrinsic to the theory are qualitatively
discussed: (1) some limitative bounds on the mechanism accelerating
cosmic rays; (2) the degree of isotropy implied by the residence
time of the cosmic rays in the Galaxy.
\vspace{-0.5cm}
\end{abstract}

\maketitle


\section{Introduction}

A solution to the knee and ankle problem has been recently proposed
[1-4] with a major characteristic feature consisting of a
fundamental connection between the knee and the ankle mainly based
on some data on the Milky Way (size, magnetic field, gas density)
and nuclear interaction lengths of cosmic ions in the interstellar
medium. The knees and ankles of individual ions are generated by
galactic cosmic rays, though a tiny extragalactic component might be
admitted. The upper energy boundary of $5\times10^{19}$ eV marks the
iron ankle [2] in the energy region of the quasi rectilinear
propagation of iron nuclei through the Galaxy.

The present study assumes that this solution does exist as reported elsewhere [1-4] and it explores some consequences regarding:
(1) the chemical composition of the cosmic radiation above the knee up to $5\times10^{19}$ eV;
(2) some basic constraints on the acceleration mechanism in the Milky Way;
(3) the residence time of galactic cosmic rays and its relevance pertaining to the level of isotropy in the arrival directions of the cosmic rays at Earth.

No new elements of the theory are introduced in this study.

Other salient consequences of the common origin of the knee and the ankle are not discussed due to the paper size limitation. For example, the existence of the second knee around $6 \times 10^{17}$ eV [5,6] is an intrinsic property of the energy spectra resulting from this theory. A second example is that the iron knee is less pronounced than the knees of the lighter ions, in particular those of helium and proton. This circumstance  may be quite fruitful while searching for the iron knee in the energy region $10^{17}$-$10^{18}$ eV as aimed at by the Kascade-Grande experiment, presently in data taking [7].

The quantitative agreement of this theory with the experimental data of Akeno [5], Haverah Park [8] and Auger [9] is illustrated in figure 1 (other data are in fig.11 and 12 of ref.2). Note that in figure 1 only one energy point at $10^{14}$ eV is used to normalize the predicted cosmic-ray intensity.


\begin{figure}[htb]
\vspace{-0.3cm}
\includegraphics [angle=0,width=8cm,height=8cm] {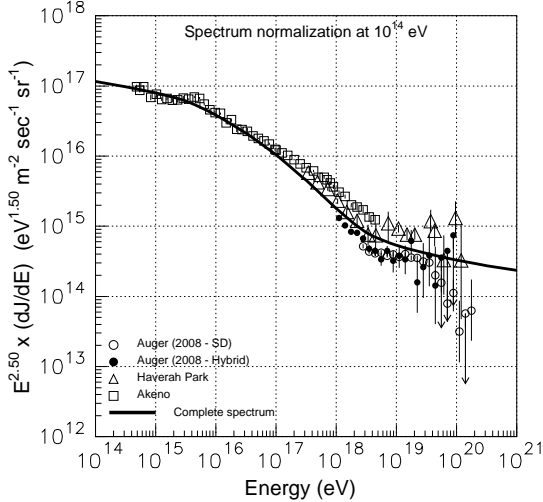}
\vspace{-2cm}
\caption{The thick line is the result of calculation with the differential cosmic ray intensity normalized to $1.16 \times 10^{-18}$ particles/($m^2$ sr s eV) at $10^{14}$ eV. The data are from the Akeno [5], Haverah Park [8] and Auger [9] experiments.}
\label{fig:largenenough}
\vspace{-0.3cm}
\end{figure}

The energy spectra of individual ions or group of elements are also in agreement with the experimental data as reported elsewhere (ref.4, fig.5).


\begin{figure}[htb]
\vspace{-0.3cm}
\includegraphics [angle=0,width=8cm,height=8cm] {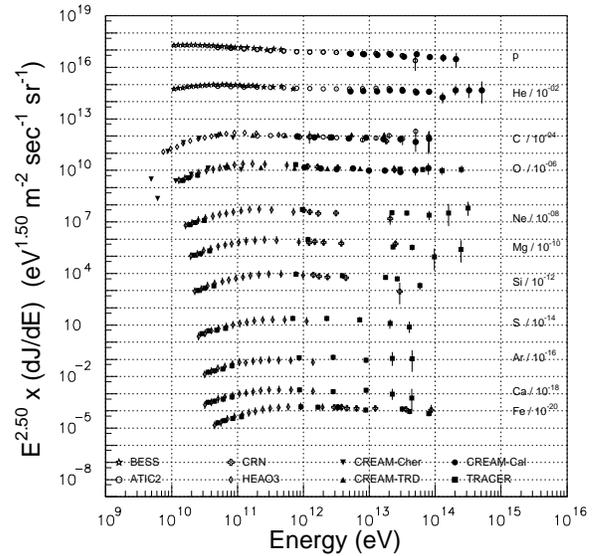}
\vspace{-1.5cm}
\caption{Energy spectra of $11$ individual ions measured by balloon-borne and satellite experiments suggesting and founding the  postulate of the constant spectral indices.}
\label{fig:largenenough}
\vspace{-0.4cm}
\end{figure}

Notice that the agreement of the computed spectrum with data displayed in figure 1 does not critically depend on the ion abundances nor on the specific normalization energy
\footnotemark
. The accord
persists, whatever the input parameters, provided that they are comprised within the observational limits defining the Galaxy and those defining nuclear interaction cross sections.

\footnotetext {Michelangelo Ambrosio and Carla Aramo from Universit\`a di Napoli, Italia, suggested to us to normalize the differential energy spectrum  at any energy in the interval $10^{11}$-$10^{14}$ eV where numerous balloon and satellite data are available. This normalization choice, unlike that at $10^{16}$ and $10^{19}$ eV [2], besides its superior reliability and precision, makes more impressive the agreement between theory and data due to the larger excursion between the normalization energy ($10^{12}$ eV) and the extreme end of the predicted spectrum ($5\times 10^{19}$ eV).}


\section{Basic elements of this theory related to the present calculation}

A logical partition of this theory would identify two areas: (1) the simulation of cosmic-ray trajectories in the Galaxy, detailed in the quoted papers [1-4] and others [10-12]; (2) the introduction of a new postulate, a fundamental part of the theory, which states that the spectral indices of all ions have constant, common values of about $2.65$ up to $5 \times 10^{19}$ eV.

The postulate is well anchored to numerous observations gathered by balloon and satellite experiments in the energy interval $10^{11}$-$10^{15}$ eV. Figure 2 shows the energy spectra of individual ions: proton, helium, carbon, oxygen, neon, magnesiun, silicon, sulfur, argon, calcium and iron. The data plainly indicate that all spectra are straight lines in logarithmic scales and that the indices accumulate around a common value of about $2.65$ as shown in figure 3.


\begin{figure}[htb]
\vspace{-0.35cm}
\includegraphics [angle=0,width=8cm] {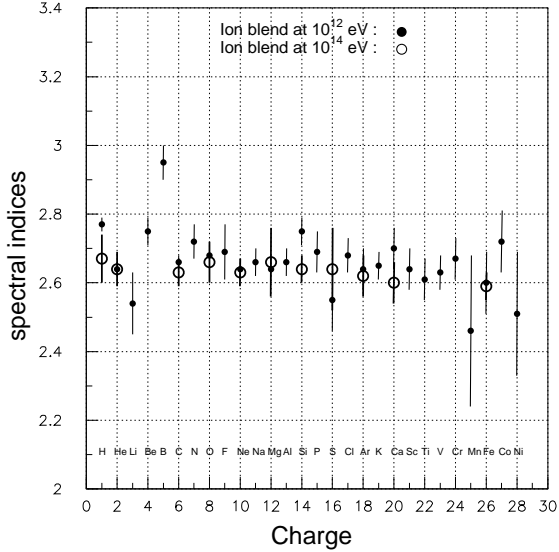}
\vspace{-1.5cm}
\caption{Spectral indices of cosmic ions from Hydrogen to Nickel measured by balloon-borne and satellite experiments at $10^{12}$ eV (filled dots) and $10^{14}$ eV (open dots). All indices extracted from the spectra around $10^{14}$ eV fall between $2.6$ and $2.7$.}
\label{fig:largenenough}
\vspace{-0.8cm}
\end{figure}

Even more important than balloon data is the Fe spectrum measured by the Kascade Collaboration [13] in the attempt to detect a signal of the yet unobserved Fe knee. In the large energy interval $10^{15}$-$8 \times 10^{16}$ eV, a constant index close to $2.64$ has been measured. Thus, the Fe spectrum measured by Kascade enlarges to $6$ energy decades the interval where the values of the spectral indices are constant and almost equal. This fundamental observation of the Kascade Collaboration reinforces the basis of the postulate.

The intensity of the computed cosmic radiation can be normalized to experimental data at any arbitrary energy in the interval $10^{11}$-$5 \times 10^{19}$ eV.


\begin{figure}[htb]
\vspace{-1.05cm}
\includegraphics [angle=0,width=9cm]{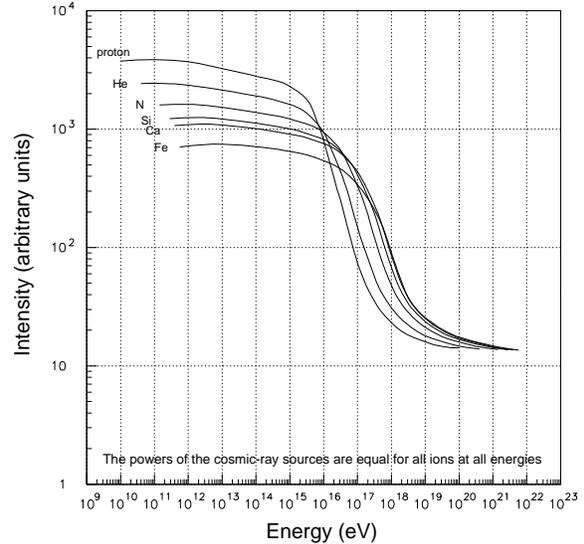}
\vspace{-1.9cm} \caption{ Relative intensities of cosmic rays
reaching the local galactic zone computed by assuming a uniform
spatial distribution of the sources in the disc volume, equal
abundances of the ions and (artificial) flat spectral indices. These
last two constraints simplify the analysis at an intermediate step;
they are replaced by the correct empirical parameters termed $ion$
$blends$, specifying ion abundances and spectral indices. These
curves represent the end product of the simulation of the cosmic-ray
trajectories in the Galaxy.} \label{fig:largenenough}
\vspace{-0.7cm}
\end{figure}

The set of parameters associated to the ion groups is concisely termed {\it ion blend}. The present study adopts two ion blends: one compiled by Wiebel-Sooth  [14] at $10^{12}$ eV (LE blend, for Low Energy) and an analogous compilation [15] made by us at $10^{14}$ eV (HE blend, for High Energy) which includes recent balloon experiments at high energy [16-17]. Table 1 reports the parameters of these two ion blends. The relevant differences between HE and LE blends reside, respectively, in the proton index, $2.67$ instead of $2.77$ and a larger proton fraction, $0.42$ instead of $0.34$. A paucity of elements heavier than helium by about 1-2 percent characterizes the LE blend compared to the HE blend.


\section{The chemical composition of the cosmic radiation above $10^{17}$ eV }

Injecting uniformly in the galactic disk the same amounts of ions at given energies in the interval $10^{11}$-$5 \, 10^{19}$ eV, the ion intensities in the local galactic zone (the Solar system) take the forms reported in figure 4. These spectra are the result of the first area of the theory in the very schematic partition mentioned in  Section 2.


\begin{figure}[htb]
\vspace{-0.3cm}
\includegraphics [angle=0,width=8cm]{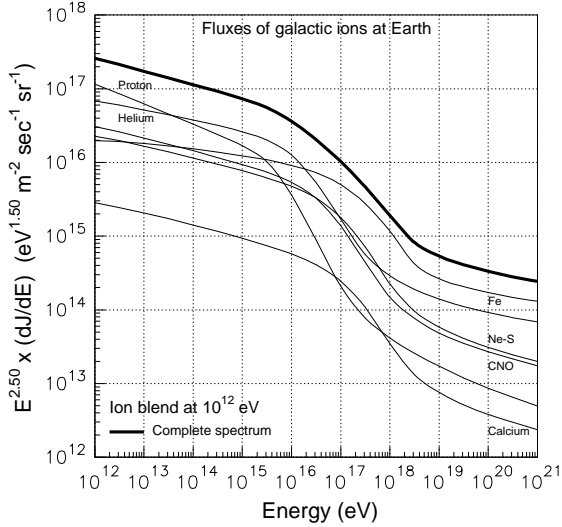}
\vspace{-2.1cm}
\caption{Computed ion spectra for the LE ion blend ($10^{12}$ eV). The thick line is the complete spectrum.}
\label{fig:toosmall}
\vspace{-0.6cm}
\end{figure}

Any features of these spectra are thoroughly determined by the theory: the difference
between the high plateau and the low plateau (see Section 9 in ref.1), the hierarchy
in the ion intensities which is controlled by nuclear cross sections and the energy regions
where the steep descents of the intensities (the knees) occur. The low-energy sides
of the energy regions where cosmic ions proceed almost unbent in the Galaxy correspond
to the regime of quasi rectilinear propagation and they mark the ankles of individual ions.

Figure 4 shows that the ion propagation through the Galaxy leaves almost unchanged the
indices defined at the cosmic-ray sources. The variations amount to $0.06$ for protons
and to $0.04$ for Fe nuclei in the interval $10^{13}$-$10^{15}$ eV.

When the spectral indices and ion abundances are included in the calculation, the regular,
smooth spectra of figure 4 spread in intensity as displayed in figures 5 and 6 for the two
ion blends. The upper curve in figure 6 (and also in figure 5) is the
{\it complete spectrum} resulting from the sum of {\it partial spectra} of individual ions or groups of ions.


\begin{figure}[htb]
\vspace{-0.3cm}
\includegraphics [angle=0,width=8cm]{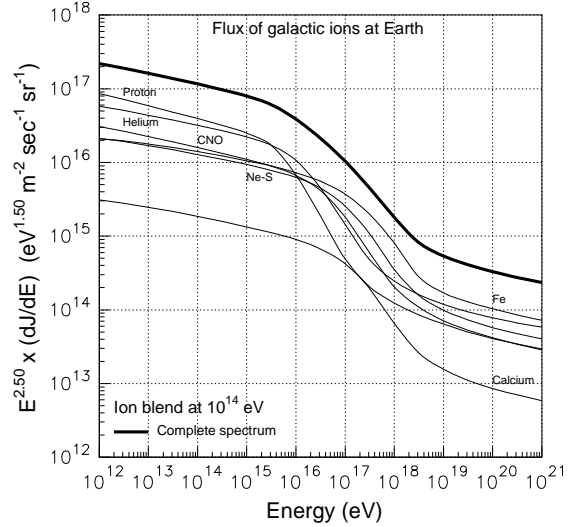}
\vspace{-2.1cm}
\caption{Computed ion spectra for the HE ion blend ($10^{14}$ eV). The thick line is the complete spectrum.}
\label{fig:largenenough}
\vspace{-1.0cm}
\end{figure}

The features of the energy spectra of two extreme ions, proton and Fe, are similar to those of the intermediate ions and, accordingly, the subsequent discussion is limited to protons and Fe nuclei. Consider for example the partial spectra of figure 6 (HE blend). The steep descent of the proton spectrum below $10^{17}$ eV (the proton knee) caused by galactic phenomena is transformed above $10^{18}$ eV into a much harder spectrum controlled only by the observed index of $2.67$ of Table 1. The Fe spectrum has the same behaviour but it extends, approximately, to an energy decade higher than that of the proton.


\begin{figure}[htb]
\vspace{-0.3cm}
\includegraphics [angle=0,width=8cm] {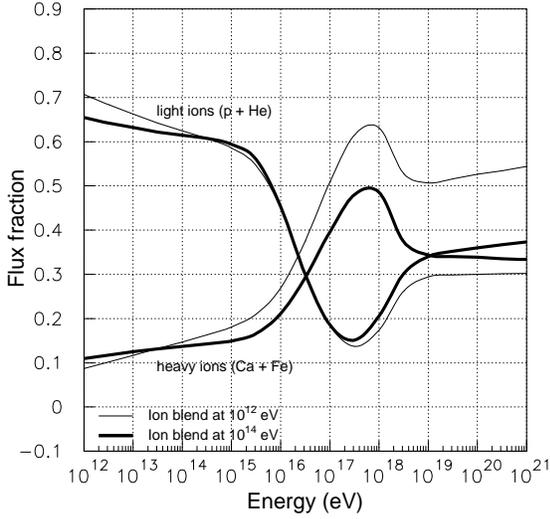}
\vspace{-2.1cm}
\caption{Relative amounts of light and heavy ions versus energy for the LE (thin line) and HE (thick line) ion blends. Notice the minimum of the light component around $3 \times 10^{17}$ eV in both ion blends, followed by the lightening of the chemical composition.}
\label{fig:largenenough}
\vspace{-0.5cm}
\end{figure}

The intensity gaps between the high and low plateaux in the proton and Fe spectra of figure 6, clearly visible in figure 4, dominate the  features of the chemical composition in this energy region. Because of these characteristics, there is an evident increase of the proton fraction relative to the Fe fraction above $3 \times 10^{17}$ eV. This trait of the energy spectra becomes evident when the relative amounts of light and heavy ions are isolated as displayed in figure 7. The light component (proton and He) has a minimum around $3 \times 10^{17}$ eV, then it increases regaining its status of a large component of about $34$ percent above $10^{19}$ eV. Above $3 \times 10^{17}$ eV a moderate lightening of the cosmic-ray spectrum takes place with a chemical composition dominated by a heavy component. A similar behaviour is exhibited by the spectra of figure 5 (LE blend).


\begin{figure}[htb]
\vspace{-0.3cm}
\includegraphics [angle=0,width=8cm]{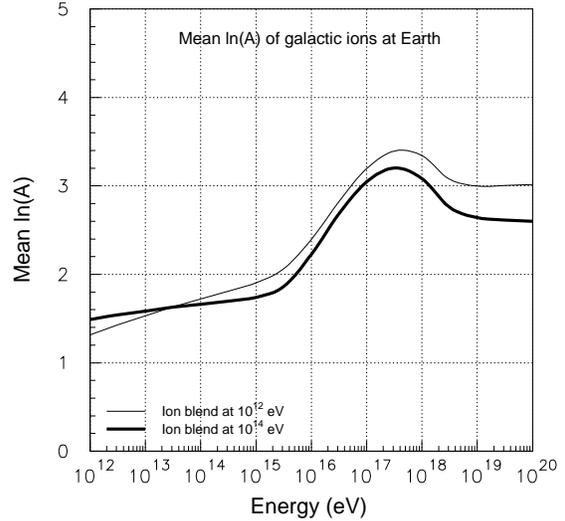}
\vspace{-2.1cm}
\caption{Mean logarithmic mass of the cosmic radiation resulting from this theory of the knee and ankle in the energy interval $10^{11}$-$5\times 10^{19}$ eV for both ion blends.}
\label{fig:largenenough}
\vspace{-0.5cm}
\end{figure}

Above $10^{19}$ eV the proton and He fraction of the LE blend (light ions in figure 7) is 30 percent while the Fe and Ca fraction (heavy ions in figure 7) is about 50 percent at $10^{19}$ eV. The light and the heavy fractions of the HE blend become almost equal at $5 \times 10^{19}$ eV amounting to $34$ percent. These significant differences in the chemical compositions resulting from the two blends are due to the softer spectral index of the proton, $2.77$, for the LE blend instead of $2.67$ for the HE blend, since the Fe indices are almost equal in both blends
\footnotemark
.
\footnotetext {A theory of particle acceleration in pulsar atmospheres predicting constant spectral indices up to $5 \times 10^{19}$ eV was conceived by Guido Pizzella in 1970 [19]. Some objections raised to the diffusive shock acceleration mechanism mentioned in the next Section 5 can also be applied to this theory.}


\section{Comparison of the computed $<$$ln(A)$$>$  with the experimental data}

Presently, experiments operating above $10^{17}$ eV cannot
distinguish individual ions or a restricted group of elements as
accomplished, for example, by the Kascade experiment below $0.8
\times 10^{17}$ eV. At very high energy, experiments determine the
mean logarithmic mass of the cosmic radiation, $<$$ln(A)$$>$, a
physical quantity whose measurement exploits the properties of the
giant atmospheric cascades requiring long elaborations. Around the
knee energy region the cosmic-ray composition is extracted from the
relative amounts of the muon and electron component of the
atmospheric cascades at ground (see, for example, Eas-top and Macro
data [18]). Above $10^{17}$ eV experiments measure the atmospheric
depth of the shower maximum, $X_{max}$, which is converted into
$<$$ln(A)$$>$ using a procedure described elsewhere [23a].

Figure 8 gives the mean logarithmic mass versus energy of the galactic cosmic rays resulting from the spectra shown in figures 5 and 6.


\begin{figure}[htb]
\vspace{-0.3cm}
\includegraphics [angle=0,width=8cm] {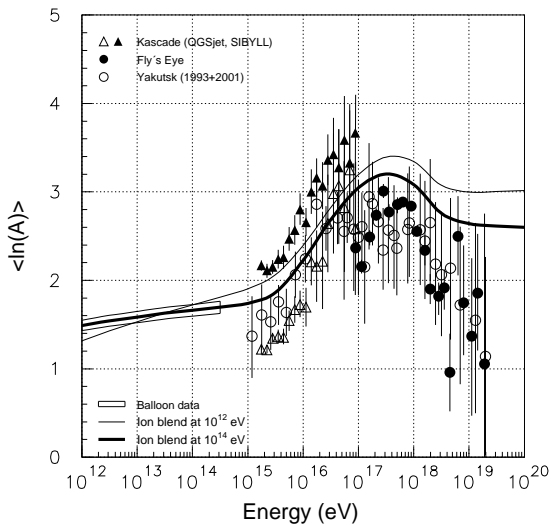}
\vspace{-2cm}
\caption{Comparison of the computed mean logarithmic mass with Kascade [13],  Yakutsk [20] and Fly{'}s Eye [21] data (selected out of many others).}
\label{fig:largenenough}
\vspace{-0.6cm}
\end{figure}

The highly distinctive feature of the computed chemical composition (i.e. $<$$ln(A)$$>$) above $3 \times 10^{17}$ eV is a definite lightening which occurs after a constant increase at lower energies, in the band $10^{15}$-$10^{17}$ eV. Above $5 \times 10^{18}$ eV the chemical composition becomes almost constant up to the maximum energy considered here, $5 \times 10^{19}$ eV.
Note that the inversion energy at $3 \times 10^{17}$ eV (HE blend), where an heavy composition becomes lighter, is rather insensitive to the ion blend ($5 \times 10^{17}$ eV for the LE blend). The characteristic energy of $3 \times 10^{17}$ eV of the maximum is not a trivial result of the theory since the four peculiarities of $<$$ln(A)$$>$,  e.g.
(1) increasing from the preknee value of $1.6$,
(2) reaching a maximum,
(3) decreasing, (4) and finally, levelling off to the value of 2.6, might have occured in another region, still preserving all of the features (1)-(4).

The comparison of the computed $<$$ln(A)$$>$ with the data of three representative experiments is made below in the range $10^{15}$-$5\times 10^{19}$ eV.


\begin{figure}[htb]
\vspace{-0.3cm}
\includegraphics [angle=0,width=8cm] {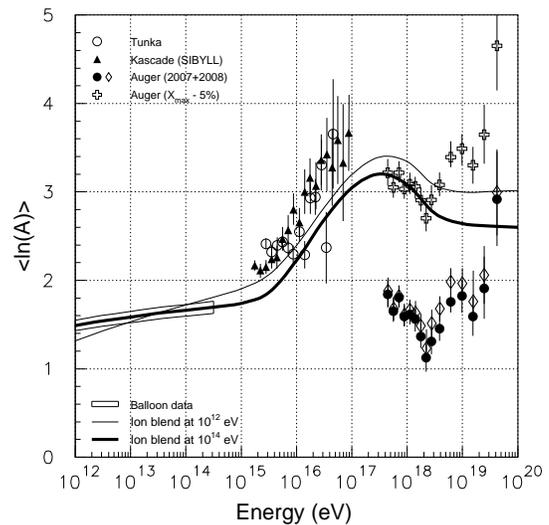}
\vspace{-2cm}
\caption{Comparison of the computed $<$$ln(A)$$>$ with the Tunka [24], Kascade [13] and Auger [23] data. A significant discrepancy between this calculation and the Auger data appears above $4 \times 10^{17}$ eV. The data represented by crosses are obtained by decreasing by $5$ percent the $X_{max}$ measured by the Auger Collaboration [26], which amounts to $35$ $g/cm^{2}$. This arbitrary operation would force an agreement with the theory in the band $4 \times 10^{17}$-$2 \times 10^{18}$ eV.}
\label{fig:largenenough}
\vspace{-0.6cm}
\end{figure}

The $<$$ln(A)$$>$ from Kascade data [13] is shown in figure 9 along with that extracted from Yakutsk [20] and Fly{'}s Eye [21] data.
Calculation and data agree exhibiting a common rise of the mean logarithmic mass from about $1.8$ up to $3.0$ in the energy band $2 \times 10^{15}$-$0.8 \times 10^{16}$ eV. Notice that both algorithms adopted  in the Kascade data analysis (QGSjet and Sibyll) exhibit a rising trend of $<$$ln(A)$$>$ versus energy. The Kascade data (open triangles) elaborated by the QGSjet algorithm at $2$-$3 \times 10^{15}$ eV are inconsistent with the extrapolated balloon data and hence excluded in the subsequent figure 10. Since individual ion spectra measured by the Kascade experiment are in accord with these calculations, the agreement on $<$$ln(A)$$>$ in figure 9 is not surprising. Notice that the Kascade ion spectra obtained by the unfolding algorithm [22], which is independent of the hadronic interaction models, give a $<$$ln(A)$$>$ in good accord with this theory.


\begin{figure}[htb]
\vspace{-0.3cm}
\includegraphics [angle=0,width=8cm] {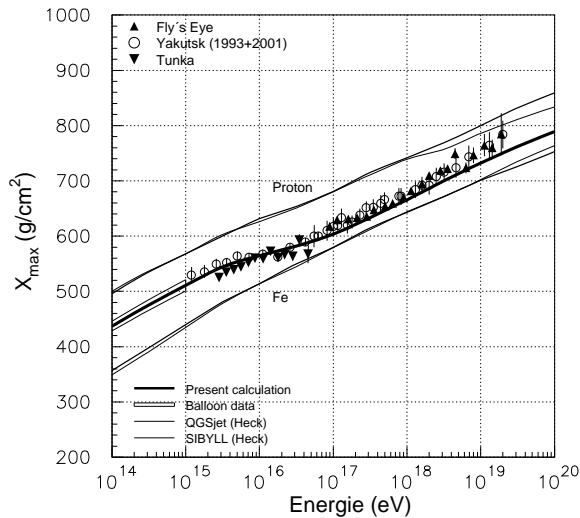}
\vspace{-2.3cm}
\caption{Atmospheric depths of shower maxima of proton and Fe nuclei versus energy. The upper and bottom double curves have been computed by Heck [25a] for two different hadronic interaction models. The $X_{max}$ resulting from this theory of the knee and the ankle (thick line inbetween) is compared with Tunka, Yakutsk and Fly{'}s Eye data.}
\label{fig:largenenough}
\vspace{-0.6cm}
\end{figure}


\begin{figure}[htb]
\vspace{-0.3cm}
\includegraphics [angle=0,width=8cm] {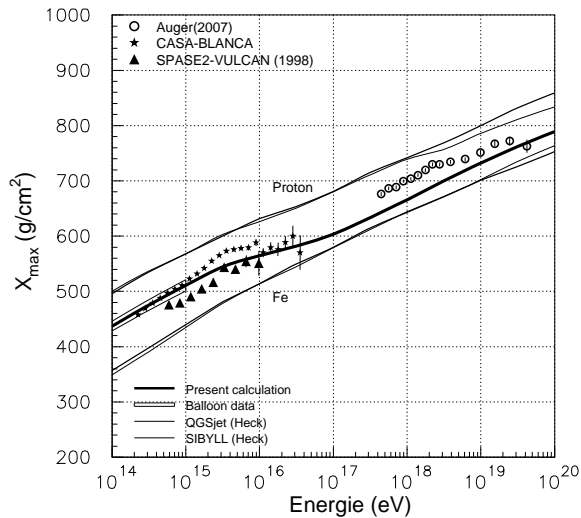}
\vspace{-2.3cm}
\caption{Atmospheric depths of shower maxima of proton and Fe nuclei versus energy. The $X_{max}$ resulting from this theory of the knee and the ankle (thick line inbetween) is compared with Auger, Spase2/Vulcan and Casa-Blanca data.}
\label{fig:largenenough}
\vspace{-0.6cm}
\end{figure}

The Yakutsk data cover both the knee and ankle energy regions. A clear rise of $<$$ln(A)$$>$ between $10^{15}$-$10^{17}$ eV is observed reaching maximum values of about $2.75$ in the band $10^{17}$-$10^{18}$ eV, then above $10^{18}$  eV, a decreasing trend dominates. Figure 9 indicates that the Yakutsk data are in accord with  those from Kascade, in the limited energy interval where data overlap, and with this theory. Fly{'}s Eye data in figure 9 show a $<$$ln(A)$$>$  thoroughly consistent with the theoretical shape reported  in figure 8 in the sense that there is an increase followed by a maximum, and finally, a decreasing trend. Unlike Yakutsk data, which have a broad maximum over two energy decades, the maximum of $<$$ln(A)$$>$ in Fly{'}s Eye data occurs in a narrower energy band, $2$-$6 \times 10^{17}$ eV, well consistent with these calculations.

A clear discrepancy exists between the computed $<$$ln(A)$$>$ and that extracted from the Auger data above $4 \times 10^{17}$ eV as shown in figure 10. A similar discrepancy also appears with the HiRes data, shown in the subsequent figure 13-a, which are well below this calculation.

In this paper all the values of $<$$ln(A)$$>$ extracted from the measured $X_{max}$ use the Heck curves shown in figure 11 [25a].

Figure 11 reports  $X_{max}$ versus energy for proton and iron with two hadronic interaction models as calculated elsewhere by others [25a]. The curves of figure 11 in the range $3 \times 10^{17}$-$5 \times 10^{19}$ eV do not substantially differ from those of the Auger Collaboration [23].

Figure 12 reports the $X_{max}$  measured by the Auger experiment [23a] corresponding to the same data of $<$$ln(A)$$>$ shown in figure 10 in order to restate the discrepancy between this calculation and the Auger data. In the low energy interval are also shown for comparison the data of Spase2/Vulcan [27] and Casa-Blanca [28].

The available data on chemical composition above $10^{17}$ eV in different experiments are not coherent and some acute inconsistencies denoted here $\alpha$, $\beta$, $\gamma$, $\delta$, $\epsilon$ and $\zeta$ emerge:


\begin{figure}[htb]
\vspace{-0.3cm}
\includegraphics [angle=0,width=8cm] {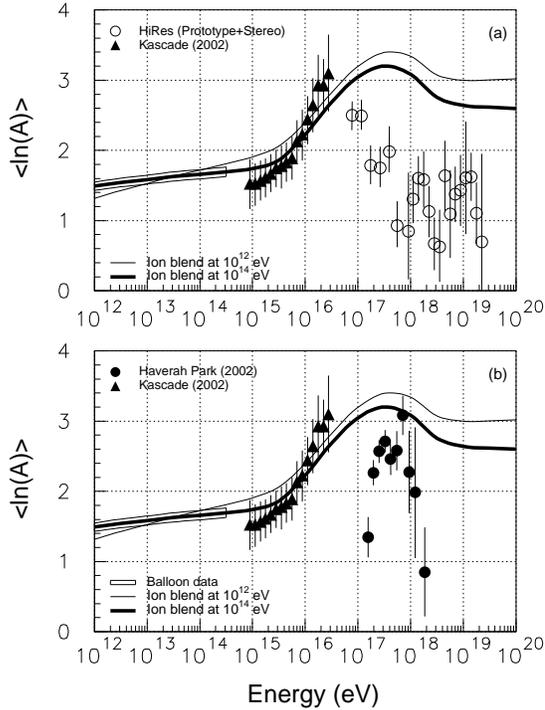}
\vspace{-2cm}
\caption{Comparison of the computed mean logarithmic mass with the Kascade [13], Hires [29] and Haverah Park [30] data. There is a large discrepancy between this calculation and the Hires data. The Haverah Park data of figure 13-b, though mimic the bell form of the computed curves, they substantially differ from the present calculation.}
\label{fig:largenenough}
\vspace{-0.3cm}
\end{figure}

($\alpha$) the  mean logarithmic mass
of about $3$-$4$ at about $0.8 \times 10^{17}$ eV determined by Kascade shown in figure 9 could not match that of $1.7$-$1.9$ inferred by us from the Auger data. The Kascade data ends at $0.88 \times 10^{17}$ eV while the Auger data initiates at $4.45 \times 10^{17}$ eV. It seems unnatural that an abrupt change in the chemical composition would take place in such a small energy band $1$-$4 \times 10^{17}$ eV. Note that $<$$ln(A)$$>$ from Kascade data is extracted using individual ion spectra and not $X_{max}$. Since an increase of $1.5$ units in $<$$ln(A)$$>$ is independent of the hadronic models used by the Kascade experiment and since balloon data have a $<$$ln(A)$$>$ of about $1.75$ in the range $10^{14}$-$10^{15}$ eV, a value of $3.2$ around $10^{17}$ is a solid, initial reference for the experiments operating above $10^{17}$ eV.

($\beta$) Auger data
on $<$$ln(A)$$>$ shown in figure 10, due to the small error bars, unambiguously separate two energy regions where the chemical composition has opposite trends with energy: below $2 \times 10^{18}$ eV decreases with energy while above $2 \times 10^{18}$ eV increases, reaching the mean value of about $3$ with a large error bar (the last data point). This rising trend badly contrasts from the opposite trend of Yakutsk and Fly{'}s Eye data in the interval $10^{18}$-$4 \times 10^{19}$ eV, and also differs from an approximate flat trend of the HiRes data in the same energy band. The focus is on the trend of $<$$ln(A)$$>$ versus energy since absolute values are difficult to be ascertained due to the inaccuracies of the hadronic interaction models which affect $X_{max}$ versus energy [26].

($\gamma$) Comparing
Yakutsk and Fly{'}s Eye data on $<$$ln(A)$$>$ at $2 \times 10^{18}$ eV (on average, about 2.2) with those of the Auger experiment ($1.12 \pm 0.15$) a large gap is registered.

($\delta$) Shifting rigidly
the Auger data on  $<$$ln(A)$$>$ in such a way to join the corresponding theoretical values of about $2.7$ around $2\times10^{18}$ eV of the HE blend (figure 10), the resulting $<$$ln(A)$$>$ at $4.2 \times 10^{19}$ eV would be $4.65 \pm 0.50$, a rather surprising high value implying a cosmic radiation devoid of proton and He nuclei.
If the  $<$$ln(A)$$>$ of $2.2$ around  $2\times 10^{18}$ eV, resulting from the Yakutsk and Fly{'}s Eye experiments, is used in the shift (instead of the theoretical value of 2.7), the Auger data at $4.2 \times 10^{19}$ eV would still have an ultraheavy chemical composition.
This bland form of {\it reductio ad absurdum} ventilates that the observed $X_{max}$ versus energy of the Auger experiment not only would necessitate a rigid shift but also a substantial deformation of the entire curve.

($\epsilon$) Notice
that the simulated $X_{max}$ distribution of the Auger experiment around $10^{19}$ eV has a width (FWHM) of about $104$ $g/cm^{2}$ for protons
and $44$ $g/cm^{2}$ for Fe nuclei, while the corresponding $X_{max}$ are, respectively, $750$ $g/cm^{2}$ and $680$ $g/cm^{2}$ [26b].
From a model-free analysis of the Auger Collaboration an $X_{max}$ resolution of $20$ $g/cm^{2}$ is estimated [23b]. The  uncertainty of $22\%$ [23a] in the energy scale is equivalent to about $5$ $g/cm^{2}$ taking into account the slopes of $X_{max}$ versus energy (fig.3, ref.23a). Finally, the different interaction models globally contribute to the uncertainty with about $15$ $g/cm^{2}$.

Suppose to arbitrarily and rigidly shift by $20$ $g/cm^{2}$ the $X_{max}$ versus energy observed by the Auger experiment.
Since the corresponding $<$$ln(A)$$>$ resulting from this arbitrary shift at $2 \times 10^{18}$ eV cannot decrease downwardly (in such an hypothetical situation the cosmic radiation is devoid of all nuclei heavier than He), the $<$$ln(A)$$>$ must displace upwardly in figure 10.
The $<$$ln(A)$$>$ resulting from this artificial shift would imply smaller atmospheric depths, i.e. an overabundance of heavy nuclei, which in turn entails a width of the $X_{max}$ distribution narrower than that expected from a cosmic radiation saturated by protons and He nuclei. The conclusion is that the atmospheric cascades in the present Auger data sample should exhibit a mismatch between  $\sigma (X_{max})$ and $X_{max}$.

($\zeta$) Figure 13
reports the $<$$ln(A)$$>$ extracted from the Hires data [29] and that from Haverah Park experiment with a revised data sample [30]. The conversion of the measured proton fraction $F_{p}$ (see fig.11, ref.[30]) to $<$$ln(A)$$>$ is obtained by assuming that the rest of cosmic-ray fraction, $1-F_{p}$, consists of Fe ions. The $<$$ln(A)$$>$ of the Hires experiment has a decreasing trend in the band $(1-8)\times 10^{17}$ eV while the Haverah Park experiment has the opposite trend in the same energy interval. The average difference in the values of $<$$ln(A)$$>$ in the two experiments amounts to about $1.4$ units of $<$$ln(A)$$>$ in the range $5 \times 10^{17}$-$10^{18}$ eV, a large discrepancy.


\section{Some bounds on the unknown engine accelerating galactic cosmic rays}

The engine accelerating cosmic rays in the Milky Way will release them ubiquitously with constant spectral indices, in the limited energy range explored here $10^{11}$-$5 \times 10^{19}$ eV. This is a basic tenet of this theory of the knee and the ankle. An outdated, still widespread hypothesis dictates that the bulk of cosmic rays are accelerated in supernovae remnants through the mechanism of diffusive shock acceleration. How does this hypothesis compare with the postulate of constant spectral indices ? The comparison is immediate.

Empirically, it has been demonstrated in a set of 12 supernova remnants that the maximum energies of the electrons rarely exceed $10^{13}$ eV [31]. Protons should conform to this limit.

Theoretically, particles would attain a maximum characteristic energy by the diffusive shocks acceleration in supernovae remnants due to the limited number of kicks they can acquire during the supernova remnants lifetimes. For example, in a classical paper [32] the maximum proton energy is estimated to be around $10^{14}$ eV and extended to $10^{17}$ eV in particular environments [33].


\begin{figure}[htb]
\vspace{-0.3cm}
\includegraphics [angle=0,width=8cm] {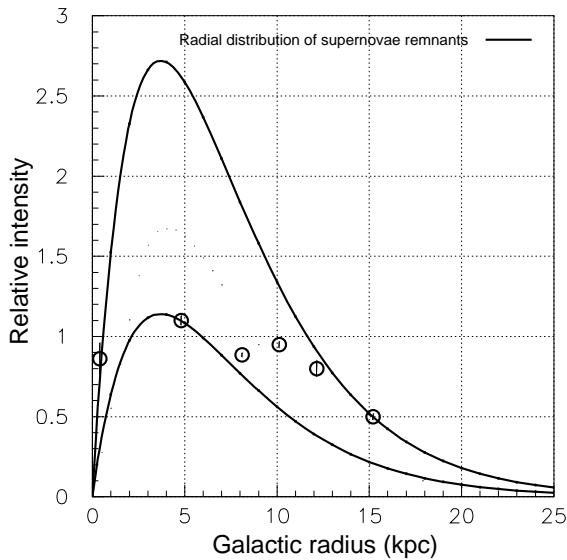}
\vspace{-1.6cm}
\caption{Radial distribution of galactic cosmic rays (open dots) [36,37] producing the $\gamma$ rays with energy greater than $100$ MeV as detected by the EGRET experiment. The radial distribution of supernovae remnants in the Milky Way (solid lines) adjusted to the function $r\, exp(-r/r_{0})$, where $r_{0}$=$3.7$ kpc [38] normalized to the cosmic-ray intensity extracted from EGRET data at $r=5$ and $r=15$ kpc.}
\label{fig:largenenough}
\vspace{-0.7cm}
\end{figure}

If hadronic cosmic rays were accelerated in supernovae remnants with a maximum limit in energy of $10^{14}$ eV (or $10^{17}$ eV), it would follow that another engine, or more engines, would have to operate the acceleration in the huge energy interval, $10^{14}$-$5 \times 10^{19}$ eV in order to comply with the postulate of constant spectral indices supported by the data as shown in figures 2 and 3. The fact that the  spectrum of cosmic rays in logarithmic scales of intensity and energy is an approximate straight line for many orders of magnitudes would imply a perfect matching of different accelerating engines, in different energy bands, active in cosmic regions of dissimilar sizes, having particle injection efficiencies tuned to each engine. But the matching of these heterogeneous hypothetical engines leading to constant indices in the range $10^{11}$-$5 \times 10^{19}$ eV appears implausible.

In the above reasoning it is silently implied that the diffusive shocks would accelerate cosmic rays below $10^{14}$ eV. The same argument applied in the low energy end of the cosmic ray spectrum, below $10^{14}$ eV, taking into account the data shown in figures 2 and 3, relegates the diffusive shock acceleration in the domain of imagination.

This inference is not isolated because a number of empirical objections, from different observational areas, constrast with the hypothesis of diffusive shock  acceleration in supernova remnants:
(1) since the total power transported by galactic cosmic rays is $10^{41}$ erg per second [11] and the average kinetic energy converted into the bulk motion of the supernovae is $10^{51}$ erg, occurring at a rate of 2-3 explosions per century, the resulting conversion efficiency for cosmic-ray acceleration is too high compared to that observed in other astrophysical environments. For example, in nova explosions the acceleration efficiency of relativistic electrons is estimated to be $1\%$ [34].
(2) Direct observations of $\gamma$ ray intensities from many supernovae remnants (for example Cas A) indicate a deficit of primary protons [35].
(3) The radial distribution of supernova remnants is not compatible with the radial distribution of cosmic rays producing the observed gamma ray spectra [36,37]. Figure 12 compares the two radial distributions normalized to the data points at the arbitrary galactocentric radii of $5$ and $15$ kpc where the common values are, respectively, $1.2$ and $0.50$ [38]. Note that the prominence of the inconsistency results from the relative high precision measurements of the two data samples. Any of the two normalizations or others (not shown) equally evince disagreement.

Note that the observational evidence that electrons of $10^{13}$ eV are detected in supernova remnants is not a label that the diffusive shock acceleration does actually accelerate them. Plausibly, an ubiquitous mechanism, still unknown and still undescribed in detail, operates in supernova remnants and in other astrophysical sites as suggested by $\gamma$ ray spectra observed in the range $10^{11}$-$10^{14}$ eV. The $\gamma$ ray spectra, having indices in the range $1.8$-$2.4$ as detected by Hess, Cangaro, Veritas and other observatories, indicate that the processes generating $\gamma$ rays in supernova remnants are the same at work in O-B star associations, pulsar atmospheres, giant molecular clouds and galactic nuclei. These findings directly indicate that cosmic-ray acceleration is pervasive.

The conclusion is that the acceleration of both hadronic cosmic rays and electrons through the diffusive shock acceleration in supernovae remnants is not compatible with the observed constant spectral indices and hence unreal
\footnotemark
.
\footnotetext {Here it is suggested that the mechanism of diffusive shock acceleration does not operate in nature the high energy acceleration of the bulk of the cosmic rays. The cosmic-ray acceleration at high energy in the magnetic clouds wandering in the Galaxy, proposed by Enrico Fermi in $1954$, has been proved not to operate in nature [39] and it constitutes another example of failure of historical importance in the quest for the true mechanism accelerating cosmic rays.}


\section{Residence time of the cosmic rays in the Galaxy and anisotropy}

The explanation of the knee and ankle [1-4] implies a residence time of galactic ions with highly distinctive features. The residence time is directly proportional to the grammage because at high energy all ions travel at the same velocity, the speed of light.

In the following, using qualitative arguments, it is suggested that the residence time implicit to this theory of the knee and the ankle can be consistent with the isotropy level of the cosmic radiation measured by many experiments at low energies, i.e. below $10^{17}$ eV.

A long-standing, unsolved, severe problem in Cosmic Ray Physics is the large discrepancy between the computed anisotropy level and the measured one. Particle propagation processes through the Galaxy, computed by analytical methods which assume diffusive propagation, determine the anisotropy in the arrival direction of cosmic rays at Earth. The computed anisotropy at $10^{14}$ eV [40] is an order of magnitude higher than that measured.

A similar evaluation with a different computational method has been accomplished by Hillas [41,42] adopting a residence time of the form $\tau$=$K$/$E^{\delta}$, where $K$ is a  normalization constant, $E$ the cosmic-ray energy and $\delta$ a suitable index.

From low energy nuclear fragmentation data $\delta$ was believed to be in the range $0.5$-$0.6$, while adopting some forms of theoretical turbulence (Kolmogorov and Kraichnan)
in the interstellar medium $\delta$
may decrease to $0.33$. The lower the value of $\delta$, the longer the residence time of cosmic rays in the Galaxy. Longer residence times necessarily imply higher level of isotropisation.

The residence times intrinsic to this theory of the knee and the ankle for He and Fe ions below $10^{17}$ eV have been calculated. In the range $10^{11}$-$10^{16}$ eV $\delta$ results $0.08$ [43], while in the range $10^{16}$-$10^{17}$ eV $\delta$=$0.35$. Due to the small value of $\delta$ below $10^{17}$ eV an isotropisation level higher than that resulting from the quoted calculation [42], which adopted $\delta$=$0.6$ and $\delta$=$0.33$, is expected. Hence, by this indirect and involved argument, the anisotropy expected from this theory below $10^{17}$ eV should be closer to observations than that of the quoted calculation [42].

Above $10^{17}$ eV a direct calculation has not yet been performed due to the large number of events required for the evaluation of the isotropy level.


\section{Conclusions.}

Data on $<$$ln(A)$$>$ are in accord with the present calculation about:
(1) the rising trend of $<$$ln(A)$$>$ in the band $10^{15}$ to $3 \times 10^{17}$ eV observed in many experiments.
(2) The maximum absolute value of about $3.2$ of $<$$ln(A)$$>$ around $3 \times 10^{17}$ eV at the confluence of Kascade [13], Yakutsk [20] and Fly{'}s Eye [21] data.
(3) The lightening of $<$$ln(A)$$>$ in the interval $3 \times10^{17}$ to $4 \times 10^{18}$ eV.
(4) The particular energy of $3 \times 10^{17}$ eV where the heavy cosmic-ray composition initiates a characteristic decline.
(5) The relative lightening of the cosmic radiation in the band $4 \times 10^{17}$ to $2 \times 10^{18}$ measured by the Auger experiment, which is in accord with this calculation, though its absolute value is incompatible (fig.10).

The Hires data shown in figure 13-a completely disagree with the
present calculation based on spectral indices of Table 1.

The fourfold circumstance 1, 2, 3 and 4 mentioned above and the
existence of the second knee (see figures 5 and 6) constitute an
unmistakable multiple fingerprint of this theory anchored to the
available experimental data.

The failure of the present calculation above $10^{18}$ eV, explicit
in figures 10 and 13-b, does not affect the theory of the knee and
the ankle [1-4] but only the ion indices of Table 1. The Fe index of
$2.64$ as measured by the Kascade Collaboration, instead of 2.62 of
Table 1, and similar changes at less than $1\%$ level, suffice to
reconcile the theory with the experimental data on $<$$ln(A)$$>$
above $10^{18}$ eV still preserving the accord below this energy.

The light chemical composition resulting from the Auger experiment
above $10^{18}$ eV, due to its superior precision compared to other
experiments, if confirmed and assessed, would signal a new
fascinating aspect of the cosmic radiation, absolutely simple and
elegant: the abundances of cosmic ions at the sources observed at
low energies remain unchanged up to very high energies, $5 \times
10^{19}$ eV. The upheaval in the ion abundances registered by all
experiments between $10^{15}$ and $10^{18}$ eV is caused by some
galactic characteristics which, surprisingly, tolerate the pervasive
mechanism accelerating cosmic ions in the Milky Way. This image of
the $\it Nature$ is a vivid echo of the empirical evidence that the
spectral index of cosmic radiation of $2.7$ below $10^{15}$ eV
persists unaltered in the huge energy interval $4\times 10^{18}$ to
$5 \times 10^{19}$ eV.

\vspace{-0.2 cm}

\begin{table}[htb]
\begin{center}
\caption{ Parameters of this theory of the knee and ankle termed Low Energy (LE) and High Energy (HE) ion blends. Intensities are $E^{2.5}$ $\times$ $flux$ in units of $(m^{-2} sr^{-1} s^{-1} eV^{1.5})$.}

\begin{tabular}{lrrrr}
\hline
Blend.        & LE              &          & HE              &         \\
\hline
              & $10^{12}$ eV    &          & $10^{14}$ eV    &         \\
\hline
              & \%              & $\gamma$ & \%              & $\gamma$ \\
\hline
H             & 42.4            & 2.77     & 33.5            & 2.67     \\
He            & 26.5            & 2.64     & 27.3            & 2.64     \\
CNO           & 11.9            & 2.68     & 13.6            & 2.65     \\
Ne-S          &  9.2            & 2.67     & 10.9            & 2.63     \\
Ca(17-20)     &  1.2            & 2.67     &  2.8            & 2.63     \\
Fe(21-28)     &  8.7            & 2.59     & 12.0            & 2.62     \\
\hline
              & Flux            & $\gamma$ & Flux            & $\gamma$ \\
\hline
H             & $1.15\ 10^{17}$ & 2.77     & $3.93\ 10^{16}$ & 2.67     \\
He            & $7.19\ 10^{16}$ & 2.64     & $3.20\ 10^{16}$ & 2.64     \\
CNO           & $3.24\ 10^{16}$ & 2.68     & $1.60\ 10^{16}$ & 2.65     \\
Ne-S          & $2.50\ 10^{16}$ & 2.67     & $1.28\ 10^{16}$ & 2.63     \\
Ca(17-20)     & $3.14\ 10^{15}$ & 2.67     & $3.27\ 10^{15}$ & 2.63     \\
Fe(21-28)     & $2.36\ 10^{16}$ & 2.59     & $1.41\ 10^{16}$ & 2.62     \\
\hline
Total         & $2.71\ 10^{17}$ & 2.70     & $1.18\ 10^{17}$ & 2.64     \\
\hline
\end{tabular}
\end{center}
\vspace{-1.5 cm}
\end{table}


\end{document}